\documentclass[12pt]{article}

\usepackage{amsmath, amssymb, slashed}
\usepackage[dvips]{graphicx}
\usepackage{epsf}
\usepackage{color}

\begin{document}
  
\begin{titlepage}
\begin{center}

\hfill IPMU-11-0110 \\
\hfill \today

\vspace{1.5cm}
{\large\bf Testing Little Higgs Mechanism at Future Colliders}
\vspace{2.0cm}

{\bf Keisuke Harigaya}$^{(a)}$, 
{\bf Shigeki Matsumoto}$^{(a)}$,
\\
{\bf Mihoko M. Nojiri}$^{(b,a)}$
and 
{\bf Kohsaku Tobioka}$^{(a)}$

\vspace{1.0cm}
{\it
$^{(a)}${IPMU, TODIAS, University of Tokyo, Kashiwa, 277-8583, Japan} \\
$^{(b)}${\it Theory Group, KEK, Tsukuba, 305-0801, Japan}
}
\vspace{2.0cm}

\abstract{
In the framework of the little higgs scenario, coupling constants of several interactions are related to each other to guarantee the stability of the higgs boson mass at one-loop level. This relation is called the little higgs mechanism. We discuss how accurately the relation can be tested at future $e^+e^-$ colliders, with especially focusing on the top sector of the scenario using a method of effective lagrangian. In order to test the mechanism at the top sector, it is important to measure the Yukawa coupling of the top partner. We consider higgs associated production and threshold production of the top partner, and find that the mechanism can be tested precisely using the associate production when the center of mass energy is large enough. The threshold production also allows us to test it even if the center mass energy is not so large.}

\end{center}
\end{titlepage}
\setcounter{footnote}{0}

\section{Introduction}

It is well known that the standard model (SM) has a serious problem called the little hierarchy problem~\cite{little hierarchy}, which is essentially from quadratically divergent corrections to the higgs mass term. The little higgs scenario~\cite{LH} has been proposed to solve the problem. In the scenario, the higgs boson is regarded as a pseudo Nambu-Goldstone boson associated with a spontaneous breaking of a global symmetry at the energy scale of ${\cal O}(10)$ TeV. Explicit breaking terms of the symmetry are also specially arranged to cancel the quadratically divergent corrections at one-loop level. The mechanism of this cancellation is sometimes called the little higgs mechanism, which is commonly equipped in all models of the little higgs scenario.

The little higgs mechanism predicts the existence of new particles at
the scale of ${\cal O}(1)$ TeV, which are called little higgs
partners. The mechanism also predicts some relations between coupling
constants of SM interactions and those of the new particles. Among the
partners, the top partner is the most important one, because it is
responsible for the cancelation of the largest quadratically divergent
correction to the higgs mass term. The top partner has a color charge
and could be produced in near future~\cite{LHatLHC} at the large hadron
collider (LHC)~\cite{LHC}. The discovery of the top partner, however,
does not mean the confirmation of the little higgs scenario, because new
particles which are similar to the top partner are also predicted in
various new physics extensions of the SM\cite{tprime}. In order to test the little higgs scenario, we have to verify the relation between interactions predicted by the little higgs mechanism.

This verification requires us to measure the Yukawa coupling of the top
partner. Future linear colliders such as the international linear
collider (ILC)~\cite{ILC} and the compact linear collider (CLIC)~\cite{CLIC} give a good opportunity for coupling measurements~\cite{LHatILC}. Following four processes are considered in this article; higgs associated productions ($e^+e^- \to T \bar{T} h$, $t \bar{T} h + T \bar{t} h$) and threshold productions ($e^+e^- \to T \bar{T}$, $T\bar{t} + t\bar{T}$), where $t$, $T$, and $h$ are top quark, top partner, and higgs boson, respectively. We found that the coupling can be measured precisely using the associate production $e^+e^- \to T \bar{T} h$ when the center of mass energy is large enough. The threshold production $e^+e^- \to T \bar{T}$ also allows us to measure it with the same precision. Interestingly, with smaller center of mass energy, it is even possible to measure the coupling using the threshold production $e^+e^- \to \bar{t} T + T \bar{t}$.

This article is organized as follows. In the next section, we introduce the effective lagrangian to describe the top sector of the little higgs scenario. The little higgs mechanism is quantitatively described using the lagrangian. In sections \ref{sec: associate} and \ref{sec: threshold}, higgs associated and threshold productions of the top partner are discussed with particularly focusing on how the cross sections of these processes are sensitive to the Yukawa coupling of the top partner. In section \ref{sec: detectability}, we consider how accurately the Yukawa coupling can be measured using the processes discussed in previous sections, and discuss the capability of future linear colliders to test the little higgs mechanism. Section \ref{sec: summary} is devoted to summary of our discussions.

\section{Top sector of the little higgs scenario}
\label{sec: top sector of LH}

The method using an effective lagrangian is adopted to investigate the top sector of the little higgs scenario. In the following, we introduce the lagrangian and discuss what kinds of interactions are predicted from it. We finally mention our strategy to test the little higgs mechanism at future linear colliders such as the ILC.

\subsection{Effective lagrangian}

In the little higgs scenario, the vector-like quark called the top
partner is necessarily introduced, which is responsible for the
cancellation of quadratically divergent corrections to higgs mass term
from the top quark \cite{LH}. Interactions between higgs boson, top
quark, and top partner are described by the effective
lagrangian,\footnote{For simplicity, we have omitted to write gauge
interactions of third generation quarks and top partners in the
effective lagrangian. See appendix~\ref{app: derivation} for the
derivation of the lagrangian and its correspondence to
specific little higgs models.}
\begin{eqnarray}
{\cal L}_{\rm eff}
&=&
- m_U \bar{U}_L U_R
- y_3 \bar{Q}_{3L} H^c u_{3R}
- y_U \bar{Q}_{3L} H^c U_R
\nonumber\\
&&
- (\lambda /m_U) \bar{U}_L u_{3R} |H|^2
- (\lambda'/m_U) \bar{U}_L U_R |H|^2 + h.c.,
\label{eq: effective lagrangian}
\end{eqnarray}
where $Q_{3L}=(u_{3L},b_{3L})^T$ and $u_{3R}$ are third generation left- and right-handed quarks, while $U_L$ and $U_R$ are left- and right-handed top partners. Higgs boson is denoted by $H^c$, where the superscript '$c$' denotes charge conjugation. Quantum numbers of these fields are shown in Table \ref{tab: gauge charges}. Here, we postulate that top partners couple only to third generation quarks to avoid flavor changing processes. Model parameters $m_U$, $y_3$, $y_U$ and $\lambda$ are taken to be real by appropriate redefinitions of the fields. On the other hand, the parameter $\lambda^\prime$ can be complex in general. We take, however, this parameter to also be real because of the little higgs mechanism discussed below.

\begin{table}[t]
\centering
\begin{tabular}{|c|ccccc|}
\hline
& $Q_{3L}$ & $u_{3R}$ & $U_L$ & $U_R$ & $H$ \\
\hline
SU(3)$_c$ & {\bf 3} & {\bf 3} & {\bf 3} & {\bf 3} & {\bf 1} \\
SU(2)$_L$ & {\bf 2} & {\bf 1} & {\bf 1} & {\bf 1} & {\bf 2} \\
 U(1)$_Y$ & 1/6     & 2/3     & 2/3     & 2/3     & 1/2 \\
\hline
\end{tabular}
\caption{\small Quantum numbers of $Q_{3L}$, $u_{3R}$, $U_L$, $U_R$ and $H$.}
\label{tab: gauge charges}
\end{table}

The little higgs mechanism at the top sector can be quantitatively defended by using the effective lagrangian in eq.(\ref{eq: effective lagrangian}). Since quadratically divergent corrections to the higgs mass term should be cancelled with each other at 1-loop level, the following relation between the coupling constants ($y_3$, $y_U$, and $\lambda^\prime$) is required,
\begin{eqnarray}
-2\lambda'
=
y_3^2 + y_U^2,
\label{eq: LH relation}
\end{eqnarray}
which is nothing but the little higgs mechanism at the top sector. It is thus very important to confirm the relation experimentally. The purpose of this article is to clarify what kind of observation is the most efficient for this confirmation.

Once $H$ acquires the vacuum expectation value $\langle H^c \rangle = (v/\sqrt{2}, 0)$ with being $v \simeq 246$ GeV, the electroweak symmetry (${\rm SU}(2)_L\times {\rm U}(1)_Y$) is broken, and third generation quarks are mixed with top partners. Mass matrix of these particles is
\begin{eqnarray}
\begin{pmatrix} \bar{u}_{3L} & \bar{U}_L \\ \end{pmatrix}
\begin{pmatrix}
A & B \\
C & D \\
\end{pmatrix}
\begin{pmatrix} u_{3R} \\ U_R \\ \end{pmatrix}
=
\begin{pmatrix} \bar{t}_L & \bar{T}_L \\ \end{pmatrix}
\begin{pmatrix} m_t & 0 \\ 0 & m_T \\ \end{pmatrix}
\begin{pmatrix} t_R \\ T_R \\ \end{pmatrix},
\label{eq: mass matrix}
\end{eqnarray}
where $A$, $B$, $C$, and $D$ are defined by $A \equiv y_3 v/\sqrt{2}$,
$B \equiv y_U v/\sqrt{2}$, $C \equiv \lambda v^2/(2m_U)$, and $D \equiv
m_U + \lambda^\prime v^2/(2m_U)$, respectively. We call a Dirac fermion
$t$ composed of $t_L$ and $t_R$ the top quark in following
discussions. We also call $T$ the top partner which is defined by
$T_L$ and $T_R$ in the same manner as $t$. Mixing angles for left- and right-handed quarks to diagonarize the mass matrix are then defined by
\begin{eqnarray}
\begin{pmatrix} t_L \\ T_L \\ \end{pmatrix}
=
\begin{pmatrix}
{\rm cos}\theta_L & -{\rm sin}\theta_L \\
{\rm sin}\theta_L &  {\rm cos}\theta_L \\
\end{pmatrix}
\begin{pmatrix} u_{3L} \\ U_L \\ \end{pmatrix},
\qquad
\begin{pmatrix} t_R \\ T_R \\ \end{pmatrix}
=
\begin{pmatrix}
{\rm cos}\theta_R & -{\rm sin}\theta_R \\
{\rm sin}\theta_R &  {\rm cos}\theta_R \\
\end{pmatrix}
\begin{pmatrix} u_{3R} \\ U_R \\ \end{pmatrix}.
\label{eq: mixing matrix}
\end{eqnarray}
Using model parameters ($m_U$, $y_3$, $y_U$, $\lambda$, $\lambda^\prime$), which are defining the effective lagrangian, mass eigenvalues ($m_t$, $m_T$) and mixing angles ($\tan\theta_L$, $\tan\theta_R$) are
\begin{eqnarray}
m_t
&=& \sqrt{(A^2 + B^2 + C^2 + D^2 - \Delta)/2}
\simeq y_3v/\sqrt{2},
\label{eq: mass and mixing 1}
\\
m_T
&=& \sqrt{(A^2 + B^2 + C^2 + D^2 + \Delta)/2}
\simeq m_U,
\label{eq: mass and mixing 2}
\\
\tan \theta_L
&=& (\Delta + A^2 + B^2 - C^2 - D^2)/(2AC + 2BD)
\simeq y_U v/(\sqrt{2} m_U),
\label{eq: mass and mixing 3}
\\
\tan \theta_R
&=& (\Delta + A^2 - B^2 + C^2 - D^2)/(2AB + 2CD)
\simeq (y_3 y_U + \lambda)v^2/(2 m_U^2),
\label{eq: mass and mixing 4}
\end{eqnarray}
where $\Delta$ in above expressions is defined by $\Delta \equiv \{(A^2 + B^2 + C^2 + D^2)^2-4(AD - BC)^2\}^{1/2}$. Last term in each expression is the leading approximation of ${\cal O}(v/m_U)$.

\subsection{Interactions}

Here, we discuss interactions predicted by the effective lagrangian in eq.(\ref{eq: effective lagrangian}) (and eq.(\ref{eq: effective lagrangian app})). Though the effective lagrangian in eq.(\ref{eq: effective lagrangian}) is originally defined by the model parameters ($m_U$, $y_3$, $y_U$, $\lambda$, $\lambda^\prime$), we use following five parameters ($m_t$, $m_T$, $\sin\theta_L$, $\lambda$, $\lambda^\prime$) as fundamental ones defining the lagrangian in following discussions. Parameters ($m_U$, $y_3$, $y_U$, $\tan\theta_R$) are therefore given as functions of the fundamental parameters ($m_t$, $m_T$, $\sin\theta_L$, $\lambda$, $\lambda^\prime$), which are obtained numerically by solving eqs.(\ref{eq: mass and mixing 1})-(\ref{eq: mass and mixing 4}). Gauge and Yukawa interactions including $t$, $T$ and $b$ (bottom quark) are then given by
\begin{eqnarray}
{\cal L}_{\rm int}
&=&
- g_s \bar{t} \slashed{G} t - g_s \bar{T} \slashed{G} T
- \frac{2e}{3} \bar{t} \slashed{A} t - \frac{2e}{3} \bar{T} \slashed{A} T
- \frac{g c_L s_L}{2c_W}(\bar{T} \slashed{Z} P_L t + h.c.)
\nonumber \\
&&
- \frac{g}{c_W} \bar{t} \slashed{Z}
\left(-\frac{2 s_W^2}{3} + \frac{c_L^2}{2}P_L\right) t
- \frac{g}{c_W} \bar{T} \slashed{Z}
\left(-\frac{2 s_W^2}{3} + \frac{s_L^2}{2}P_L\right) T
\nonumber \\
&&
- \frac{g c_L}{\sqrt{2}} (\bar{b} \slashed{W} P_L t + h.c.)
- \frac{g s_L}{\sqrt{2}} (\bar{b} \slashed{W} P_L T + h.c.)
\nonumber\\
&&
- y_t \bar{t} t h
- y_T \bar{T} T h
- (\bar{T} [y_L P_L + y_R P_R] th + h.c.),
\label{eq: interactions}
\end{eqnarray}
where $\slashed{G} = G_\mu^a (\lambda^a/2) \gamma^\mu$, $\slashed{W} = W_\mu \gamma^\mu$, $\slashed{Z} = Z_\mu \gamma^\mu$, and $\slashed{A} = A_\mu \gamma^\mu$ are gluon, $W$ boson, $Z$ boson, and photon fields with $\lambda^a$ and $\gamma^\mu$ being Gell-Mann and gamma matrices, while $h$ denotes higgs field. Coupling constants associated with SU(3)$_c$, SU(2)$_L$, and U(1)$_{\rm EM}$ gauge interactions are denoted by $g_s$, $g$, and $e = g s_W$ with being $s_W(c_W) = \sin \theta_W (\cos \theta_W)$, where $\theta_W$ is the Weinberg angle. We have also used the notation, $s_L (c_L) = \sin \theta_L (\cos \theta_L)$. Coupling constants associated with Yukawa interactions ($y_t$, $y_T$, $y_L$, $y_R$) have complicated forms, and these are given by
\begin{eqnarray}
y_t
&=&
\frac{c_L}{\sqrt{2}} (c_R y_3 - s_R y_U)
-
\frac{s_L v}{m_U} \left(c_R \lambda - s_R \lambda'\right)
\simeq
\frac{m_t}{v},
\label{eq: yukawa couplings 1}
\\
y_T
&=&
\frac{s_L}{\sqrt{2}} (c_R y_U + s_R y_3)
+
\frac{c_L v}{m_U} \left(c_R \lambda' + s_R \lambda\right)
\simeq
\frac{m_T}{v}s_L^2+\frac{v}{m_T}c_L\lambda',
\label{eq: yukawa couplings 2}
\\
y_L
&=&
\frac{c_L}{\sqrt{2}} (c_R y_U + s_R y_3)
-
\frac{s_L v}{m_U} \left(c_R \lambda' + s_R\lambda\right)
\simeq
\frac{s_L m_T}{v},
\label{eq: yukawa couplings 3}
\\
y_R
&=&
\frac{s_L}{\sqrt{2}} (c_R y_3 - s_R y_U)
+
\frac{c_L v}{m_U} \left(c_R \lambda - s_R \lambda'\right)
\simeq
\frac{s_L m_t}{v} + \frac{\lambda v}{m_T},
\label{eq: yukawa couplings 4}
\end{eqnarray}
where it should be emphasized again that the parameters ($c_R$, $s_R$, $y_3$, $y_U$, $m_U$) are obtained as functions of the fundamental parameters ($m_t$, $m_T$, $\sin\theta_L$, $\lambda$, $\lambda^\prime$). It can be seen in eq.(\ref{eq: interactions}) that weak gauge interactions of the top partner are governed by the mixing angle $s_L$, because these interactions are originally from those of SU(2)$_L$ doublet field $Q_{3L}$ through the mixing between $Q_{3L}$ and $U_L$. On the other hand, Yukawa interactions depend not only on $s_L$ but also on other parameters.

The little higgs mechanism at the top sector is quantitatively described by the relation between coupling constants $y_3$, $y_U$, and $\lambda^\prime$, as shown in eq.(\ref{eq: LH relation}). It can be seen in eqs.(\ref{eq: mass and mixing 1})-(\ref{eq: mass and mixing 4}) that first two parameters $y_3$ and $y_U$ are almost determined by masses of top quark and top partner ($m_t$ and $m_T$), and the mixing angle $s_L$. These parameters are not difficult to be measured precisely when the top partner is discovered. On the other hand, in order to determine the last parameter $\lambda^\prime$, we have to measure the Yukawa coupling of the top partner $y_T$, as can be seen in eqs.(\ref{eq: yukawa couplings 1})-(\ref{eq: yukawa couplings 4}).

\subsection{Representative point}

We first postulate that several physical quantities ($m_T$, $s_L$, and branching fractions of the $T$ decay) have already been measured precisely. This is actually possible because $m_T$ and the branching fractions can be measured accurately by observing the pair production of top partners, while $s_L$ can be determined by the single production of the top partner~\cite{LHatLHC}. As a representative point of these observables, we take
\begin{eqnarray}
m_T = 400 {\rm GeV},
\qquad
\sin \theta_L = 0.2,
\qquad
{\rm Br}(T \rightarrow t h)/{\rm Br}(T \rightarrow b W) = 0.98,
\label{eq: representative point}
\end{eqnarray}
where the value of third observable, namely, the ratio of branching fractions, corresponds to the one which is obtained by assuming $\lambda = 0$ with keeping the relation in eq.(\ref{eq: LH relation}). Higgs mass is fixed to be $m_h = 120$ GeV. It is worth notifying that the point satisfies all phenomenological constraints and is also attractive from the viewpoint of naturalness on the little hierarchy problem. For more details, see appendix \ref{app: representative point}.

\begin{figure}[t]
\begin{center}
\includegraphics[scale=0.55]{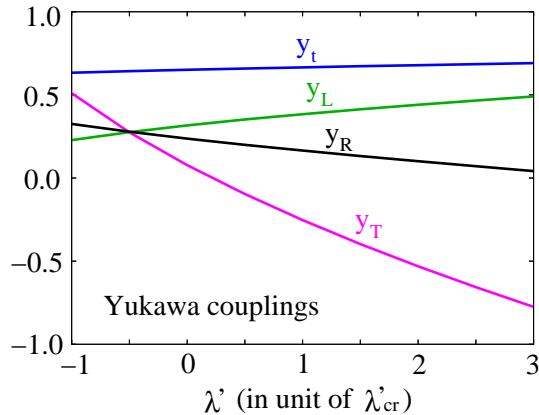}
\caption{\small Yukawa couplings ($y_t$, $y_T$, $y_L$, $y_R$) as a function of $\lambda'$ (in unit of $\lambda'_{cr}$).}
\label{fig: yukawas}
\end{center}
\end{figure}

Since the mass of top quark $m_t$ has already been measured precisely~\cite{Nakamura:2010zzi}, there are four free parameters in the effective lagrangian of eq.(\ref{eq: effective lagrangian}). It is therefore possible to test the little higgs mechanism by measuring one more observable. This observable should be sensitive to $\lambda'$, as pointed out in previous subsection. Since gauge interactions of the top partner depend only on $s_L$, we should focus on Yukawa interactions. In Fig.~\ref{fig: yukawas}, we have shown coupling constants of Yukawa interactions ($y_t$, $y_T$, $y_L$, $y_R$) as functions of $\lambda'$ in unit of $\lambda^\prime_{cr} \equiv -(y_3^2 + y_U^2)/2$, so that $\lambda' = \lambda'_{cr}$ corresponds to the prediction of the mechanism. Other model parameters are fixed to be those satisfying the conditions in eq.(\ref{eq: representative point}). It can be seen that the Yukawa coupling between top partners $y_T$ is the most sensitive against the change of $\lambda'$ as expected. As a result, we should focus on physical quantities involving this Yukawa coupling.

\section{Associate Productions}
\label{sec: associate}

It can be easily imagined that top partner productions associated with a higgs boson enable us to explore the little higgs mechanism\cite{Chang:2011jk}. It is actually possible to measure the Yukawa coupling between top partners ($y_T$) simply by measuring cross sections of the processes with an appropriate center of mass energy. There are two higgs associated processes. One is the higgs production associating with a top quark and a top partner production ($e^+ e^- \to t \bar{T} h~\&~T \bar{t} h)$, and another is the process associating with two top partners ($e^+ e^- \to T \bar{T} h)$. At the former process, the center of mass energy $\sqrt{s} \geq m_t + m_T + m_h$ is, at least, required, which corresponds to about 700 GeV in our representative point. On the other hand, at the latter process, the center of mass energy should be higher than $\sqrt{s} \geq 2m_T + m_h$, which corresponds to 920 GeV in the representative point. In this section, we consider how the cross sections of these processes are sensitive against the change of the parameter $\lambda^\prime$, and discuss which process is suitable for the confirmation of the little higgs mechanism.

\subsection{The $e^+ e^- \to t \bar{T} h + T \bar{t} h$ process}

\begin{figure}[t]
\begin{center}
\includegraphics[scale=0.54]{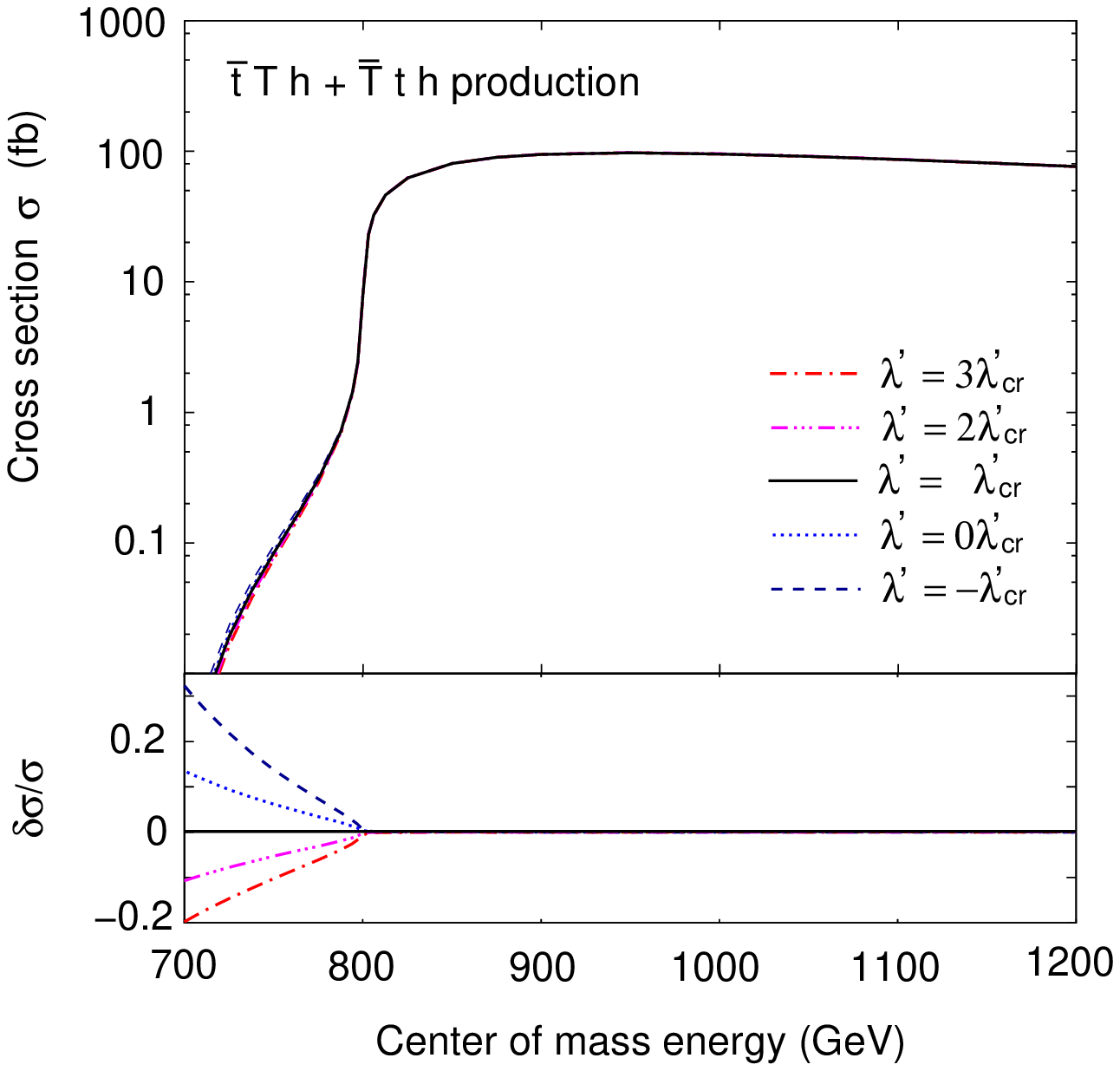}
~
\includegraphics[scale=0.54]{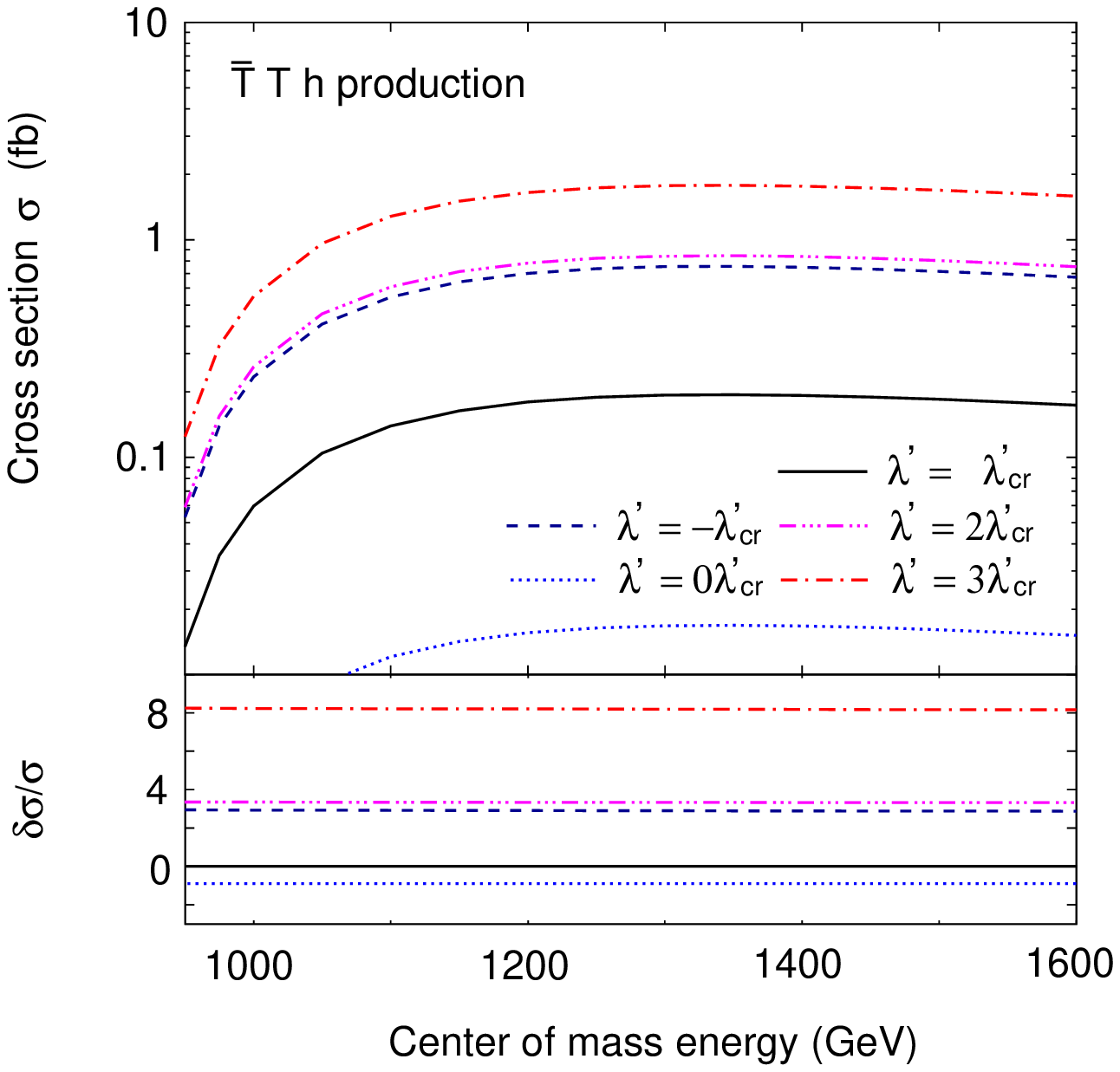}
\caption{\small Cross sections for higgs associated productions $e^+ e^- \to t \bar{T} h + T \bar{t} h$ and $e^+ e^- \to T \bar{T} h$ as a function of center of mass energy. Results for several values of $\lambda^\prime$ are shown.}
\label{fig: associate productions}
\end{center}
\end{figure}

We first consider the higgs production process associating with a top quark and a top partner. The sum of the cross sections, $\sigma(e^+ e^- \to t \bar{T} h) + \sigma(e^+ e^- \to T \bar{t} h)$, are shown in Fig.\ref{fig: associate productions} (upper part of the left panel) as a function of center of mass energy with several choices of $\lambda^\prime$. Other model parameters to depict the figure are fixed according to the representative point in eq.(\ref{eq: representative point}). In order to see the sensitivity of the cross section against the change of $\lambda^\prime$, we also plot the deviation of the cross section from the one predicted by the little higgs mechanism, namely, $\delta \sigma/\sigma \equiv [\sigma(\lambda^\prime) - \sigma(\lambda^\prime_{cr})] / \sigma(\lambda^\prime_{cr})$ (lower part of the left panel). It can be seen from the figure that the deviation becomes almost zero when the center of mass energy exceeds 800 GeV. This is because, with a center of mass energy higher than this value, two on-shell top partners can be produced, whose production cross section and branching ratio $T \to th$ are independent of the parameter $\lambda^\prime$, and this process dominates the associate production. On the other hand, the center of mass energy below 800 GeV, we can expect about 10\% deviation. The cross section is, however, quite small below 800 GeV, which is about 0.1 fb. After all, the measurement of the Yukawa coupling $\lambda^\prime$ turns out to be difficult in this associate production.

\subsection{The $e^+ e^- \to T \bar{T} h$ process}

Next, we consider the higgs production process associating with two top partners. As in the case of previous associate production, we show its production cross section $\sigma(e^+e^-\rightarrow T\bar{T}h)$ in Fig.\ref{fig: associate productions} as a function of center of mass energy with various $\lambda^\prime$ (upper part of the right panel). The deviation of the cross section from the little higgs prediction is also shown (lower part of the right panel). It can be seen from the figure that the cross section depends strongly on the value of $\lambda^\prime$, which enable us to explore the little higgs mechanism accurately using this association process, though the production cross section itself is not so large.

\section{Threshold productions}
\label{sec: threshold}

We next consider top partner productions, $e^+ e^- \to T \bar{T}$ and $ e^+ e^- \to t \bar{T} + T \bar{t}$, at the region of their threshold energies. Since the cross sections of these processes are significantly affected by the exchange of virtual higgs bosons due to the threshold singularity~\cite{Strassler:1990nw}, the Yukawa coupling $y_T$ is expected to be measured precisely. In this section, we consider how the cross sections are sensitive to the parameter $\lambda^\prime$.

\subsection{Cross section formula}

The cross section of top quark pair production at the $e^+ e^-$ collider ($e^+ e^- \to t \bar{t}$) is known to have threshold singularities due to exchanges of soft gluons and higgs bosons between top quarks~\cite{Strassler:1990nw}. Productions of top partner are, in the same manner, expected to have the same singularities, which are quantitatively obtained by using a method of non-relativistic field theory~\cite{Brambilla:2004jw}. Derivations of non-relativistic lagrangians for top partner productions and resultant cross section formulae are discussed in appendix~\ref{app: threshold productions}. For the case of top partner pair production ($e^+ e^- \to T \bar{T}$), the cross section at their threshold energy $\sqrt{s} \simeq 2 m_T$ is obtained to be
\begin{eqnarray}
\sigma_{TT}
=
\left[
\frac{16 Q_T^2 Q_e^2}{s^2}
+ \frac{12 Q_T Q_e v_T v_e}{s(s - m_Z^2)}
+ \frac{6 v_T^2 (v_e^2 + a_e^2)}{(s - m_Z)^2}
\right]
{\rm Im}\left[G_{TT}(\sqrt{s} - 2m_T; {\bf 0}, {\bf 0})\right],
\label{eq: sigma_TT}
\end{eqnarray}
where $Q_T = 2e/3$, $v_T = (g/c_W)(-2s_W^2/3 + s_L^2/4)$, $Q_e = -e$, $v_e = (g/c_W)(-1/4 + s_W^2)$, and $a_e = g/(4 c_W)$. The mass of $Z$ boson is denoted by $m_Z$. The green function $G_{TT}(E; {\bf r}, {\bf r}^\prime)$ in above formula satisfies the following Schr$\ddot{\rm o}$dinger equation,
\begin{eqnarray}
\left[-\frac{\nabla_{\bf r}^2}{m_T}
+ V_{TT}({\bf r}) - E - i\frac{\Gamma_T}{2}
\right] G_{TT}(E,{\bf r},{\bf r}^\prime)
=
\delta^3({\bf r} - {\bf r}^\prime),
\label{eq: Schrodinger TT}
\end{eqnarray}
with appropriate boundary conditions~\cite{Strassler:1990nw}. Here, $\Gamma_T$ is the total decay width of the top partner. All information of soft gluon and higgs exchanges between top partners are involved in the potential term $V_{TT}({\bf r})$, which is explicitly given by
\begin{eqnarray}
V_{TT}({\bf r})
=
-\frac{1}{|{\bf r}|}
\left( \frac{4\alpha_s}{3} + \alpha_T e^{-m_h |{\bf r}|} \right),
\label{eq: potential V_TT}
\end{eqnarray}
where $\alpha_s = g_s^2/(4\pi)$ and $\alpha_T = y_T^2/(4\pi)$. In
addition to QCD and Yukawa interactions, the electroweak interaction
(the exchange of $Z$ bosons or photons between top partners) also contributes to
the potential. This effect is however negligible and not included in our
calculations. Since one of the purposes of this article is to clarify
the effectiveness of threshold productions to measure the Yukawa
coupling $y_T$, we only consider the potential at the leading order
calculation. In order to compare theoretical predictions with
experimental results very precisely, we should include higher order
contributions especially from QCD
interactions~\cite{Brambilla:2004jw}. Since those calculations are
beyond the scope of this article and our conclusion about the
effectiveness of threshold productions is expected not to be altered
even if we include those effects\footnote{Including this effect alters
the cross section by about 10 \%.  However, the degree of deviation of
the cross section as we change $\lambda'$ is hardly affected.}, we omit the higher order contributions in our calculation.

On the other hand, for the case of top quark and top partner production, the cross section at the threshold energy region ($\sqrt{s} \simeq m_t + m_T$) turns out to be
\begin{eqnarray}
\sigma_{tT}
=
\frac{12 v_{tT}^2 (v_e^2 + a_e^2)}{(s - m_Z^2)^2}
{\rm Im}
\left[ G_{tT}(\sqrt{s} - m_t - m_T, {\bf 0}, {\bf 0}) \right],
\label{eq:sigma_TT}
\end{eqnarray}
where $v_{tT} = g c_L s_L/(4 s_W)$. The green function $G_{tT}(E; {\bf r}, {\bf r}^\prime)$ satisfies the equation,
\begin{eqnarray}
\left[-\frac{\nabla_{\bf r}^2}{2\mu_{tT}}
+ V_{tT}({\bf r}) - E - i\frac{\Gamma_{tT}}{2}
\right] G_{tT}(E,{\bf r},{\bf r}^\prime)
=
\delta^3({\bf r} - {\bf r}^\prime),
\end{eqnarray}
where $\mu_{tT}$ is the inertial mass of top quark and top partner, $\mu_{tT} = m_t m_T/(m_t + m_T)$, while $\Gamma_{tT}$ is the averaged decay width of these particles, $\Gamma_{tT} = (\Gamma_t + \Gamma_T)/2$ with $\Gamma_t$ being the total decay width of the top quark. The potential $V_{tT}({\bf r})$ is given by
\begin{eqnarray}
V_{tT}({\bf r})
=
-\frac{1}{|{\bf r}|}
\left( \frac{4\alpha_s}{3} + \frac{y_t y_T}{4\pi} e^{-m_h |{\bf r}|} \right).
\label{eq: potential VtT}
\end{eqnarray}

Finally, we would like to add a comment on the scale of the strong coupling $\alpha_s$ which appears in the potential terms $V_{TT}({\bf r})$ and $V_{TT}({\bf r})$. The scale ($\mu$) is taken to be the solution of the following self-consistency equation of the coupling,
\begin{eqnarray}
\mu = \frac{m}{2} \times \frac{4}{3}\times\alpha_s^{(run)} \left(\mu \right),
\end{eqnarray}
where $\alpha_s^{(run)}(x)$ is the running coupling of the strong interaction at the scale $x$, and $m$ is the inertial mass of a two-body system, namely, $m = m_T/2$ for top partner pair production and $m = m_t m_T/(m_t + m_T)$ for the production of top quark and top partner. This prescription is known to make higher order QCD corrections to the Coulomb potential small at the Bohr radius~\cite{Nagano:1999nw}.

\subsection{The $e^+ e^- \to T \bar{T}$ process}

\begin{figure}[t]
\begin{center}
\includegraphics[scale=0.54]{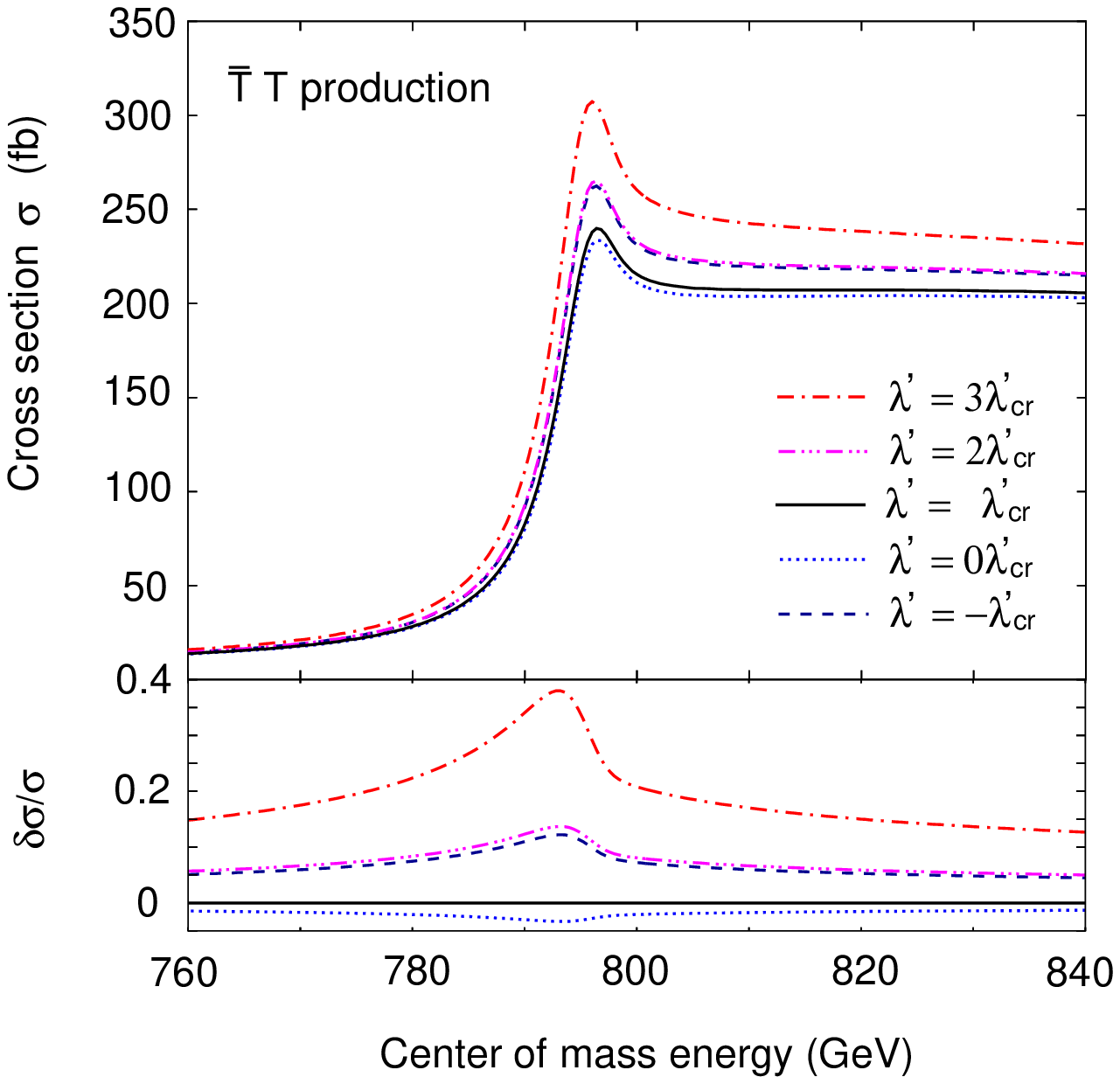}
\includegraphics[scale=0.54]{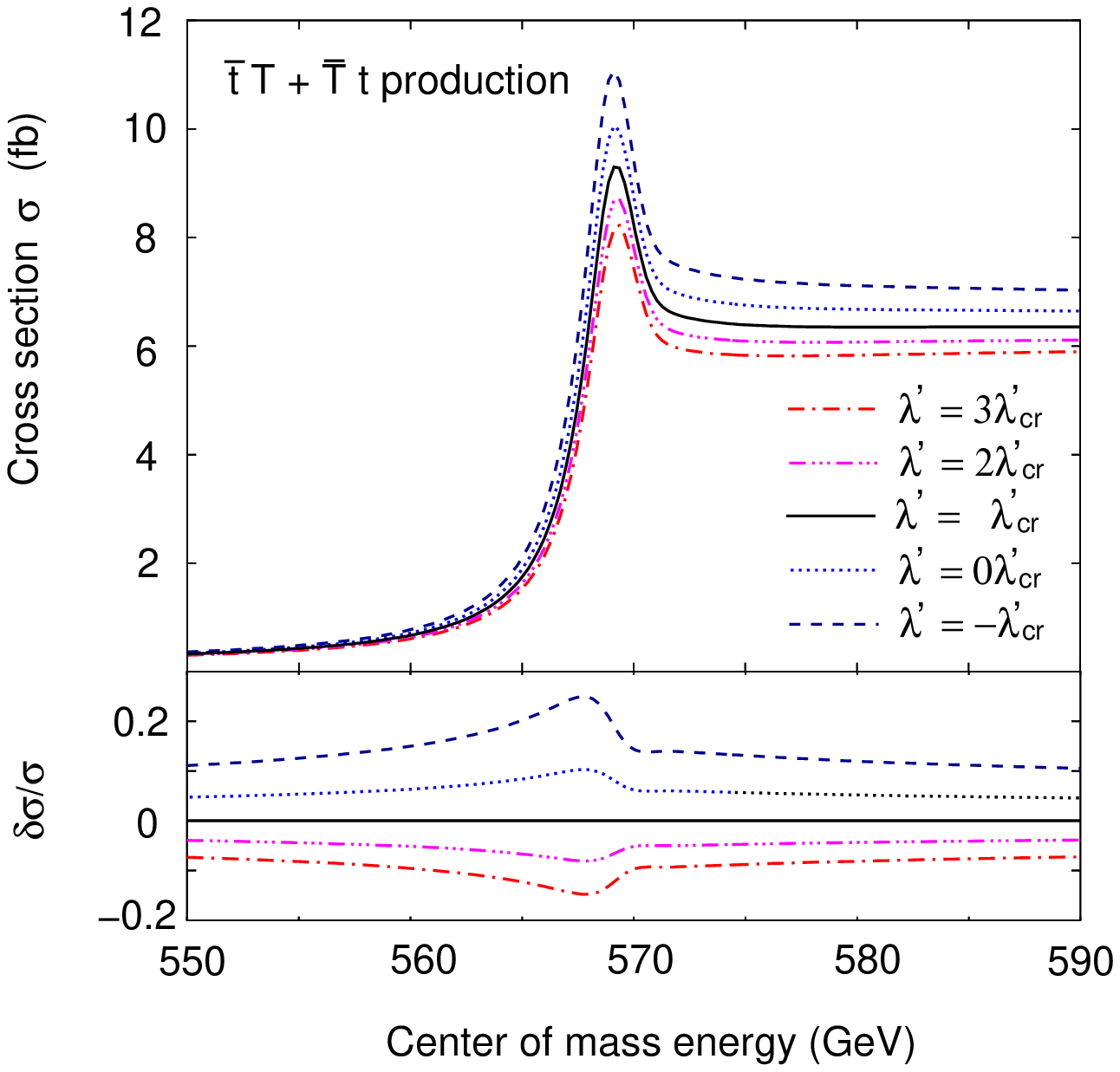}
\caption{\small Cross sections for threshold productions $e^+ e^- \to T \bar{T} $ and $e^+ e^- \to t \bar{T} + T \bar{t}$ as a function of center of mass energy. Results for several values of $\lambda^\prime$ are shown.}
\label{fig: threshold productions}
\end{center}
\end{figure}

We first consider the threshold production of top partner pair. The resultant cross section, which is obtained using the formula in eq.(\ref{eq: sigma_TT}), is shown in the left panel of Fig.~\ref{fig: threshold productions} with several choices of $\lambda^\prime$. In addition to the cross section, we also plot the deviation of the cross section from the prediction of the little higgs mechanism, $\delta \sigma/\sigma$, as in the case of higgs associated productions in previous section. It can be seen that the deviation becomes maximum at the peak of the cross section, which corresponds to the first bound state composed of top partners. Since the cross section is huge at the peak, which is about 250 fb$^{-1}$, the Yukawa coupling $y_T$ is expected to be measured precisely. It should be noticed that the potential $V_{TT}({\bf r})$ in eq.(\ref{eq: potential V_TT}) does not depend on the sign of $y_T$, so that there is two-fold ambiguity in the determination of $y_T$, as can be seen in the figure. This ambiguity is easily resolved by investigating other production channels such as higgs associated productions.

\subsection{The $e^+ e^- \to t \bar{T} + T \bar{t}$ process}

We next consider the threshold production of a top quark and a top partner. Resultant cross section and its deviation from the little higgs prediction are plotted in the right panel of Fig.~\ref{fig: threshold productions}. As in the case of top partner pair production, the deviation is again maximized at the peak corresponding to the first bound state composed of top quark and top partner. Though the cross section is lower than that of the top partner pair production, which is of the order of 10 fb$^{-1}$, it is still possible to measure $\lambda^\prime$ precisely if the integrated luminosity is large enough. One of the advantages for the use of the process to test the little higgs mechanism is that the measurement of $\lambda^\prime$ is possible even if the center of mass energy at the collision is not so large.

\section{Testing the little higgs mechanism}
\label{sec: detectability}

We are now in position to discuss the capability of future linear colliders to test the little higgs mechanism. Testing the mechanism is, as already emphasized in section \ref{sec: top sector of LH}, equivalent to the measurement of the coupling constant $\lambda^\prime$. In this section, we therefore discuss how accurately the constant $\lambda^\prime$ can be measured at collider experiments with the use of the processes discussed in previous sections.

\subsection{The $\chi^2$ function}

We focus on higgs associated production ($e^+ e^- \to T \bar{T} h$) and threshold productions ($e^+ e^- \to T \bar{T}$ and $e^+ e^- \to t \bar{T} + T \bar{t}$), because the cross sections of these processes are very sensitive to the coupling constant $\lambda^\prime$. In order to maximize the capability of collider experiments for the measurement of $\lambda^\prime$, we set the center of mass energy to be 1000 GeV, 794 GeV and 568.4 GeV for $T \bar{T} h$, $T \bar{T}$ and $t \bar{T} + T \bar{t}$ productions, respectively. On the other hand, the integrated luminosity is fixed to be ${\cal L}_{\rm eff} = 500$ fb$^{-1}$ in each process, where ${\cal L}_{\rm eff}$ is the effective luminosity defined by ${\cal L}_{\rm eff} = {\cal E} \times {\cal L}$ with ${\cal E}$ and ${\cal L}$ being the efficiency factor and the original integrated luminosity.

The efficiency factor depends on the acceptance of collider detectors and kinematical cuts used to reduce backgrounds from SM processes, which is determined precisely when experiments start. Main backgrounds against the signal processes are from SM processes of top quark production. Since cross sections of the SM processes are not too large compared to those of signal processes, background reductions are expected to be performed efficiently by imposing appropriate kinematical cuts. The efficiency factor from background reductions can be estimated using Monte-Carlo simulations. In our calculation, however, we take the efficiency factor so that the effective luminosity becomes 500 fb$^{-1}$ with simply assuming efficient background reductions. We remain the detailed calculation of the factor as a future problem.

With the use of the effective luminosity ${\cal L}_{\rm eff}$, the $\chi^2$ function, which quantifies how accurately the $\lambda^\prime$ measurement can be performed, is defined by
\begin{eqnarray}
\chi^2 (\lambda^\prime)
\equiv
\frac{\left[N(\lambda') - N(\lambda'_{cr})\right]^2}{N(\lambda'_{cr})}
=
{\cal L}_{\rm eff}
\times
\frac{\left[\sigma(\lambda') - \sigma(\lambda'_{cr})\right]^2}
{\sigma(\lambda'_{cr})},
\end{eqnarray}
where $N(\lambda^\prime) = {\cal L}_{\rm eff} \sigma(\lambda^\prime)$ is the number of the signal event with fixed $\lambda^\prime$, which will be obtained at collider experiments after imposing kinematical cuts and considering detector acceptances. Here, the value of $\lambda^\prime$ predicted by the little higgs mechanism is denoted by $\lambda^\prime_{cr} = -(y_3^2 + y_U^2)/2$. Other parameters to calculate the cross section $\sigma$ is fixed according to a representative point mentioned in section \ref{sec: top sector of LH}.

\subsection{Results}

\begin{figure}[t]
\begin{center}
\includegraphics[scale=0.6]{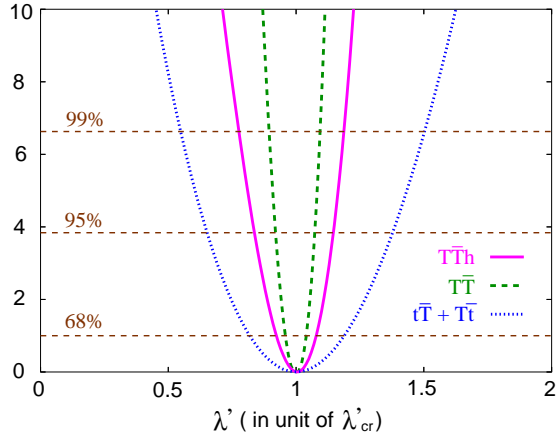}
\caption{\small The $\chi^2$ distributions for $e^+ e^- \to T \bar{T} h$, $T \bar{T}$, and $t \bar{T} + T \bar{t}$ processes.}
\label{fig: chi2}
\end{center}
\end{figure}

Resultant $\chi^2$ distributions for the higgs associated production ($T \bar{T} h$) and threshold productions ($T \bar{T}$ and $t \bar{T} + T \bar{t}$) are shown in Fig.~\ref{fig: chi2}, where the center of mass energy is fixed to be 1000 GeV, 794 GeV and 568.4 GeV, respectively. It can be seen that the coupling constant $\lambda^\prime$ can be measured with 8\% accuracy using the higgs associated production. If the center of mass energy is possible to be increased to 1350 GeV, the coupling $\lambda^\prime$ can be measured with 4\% accuracy using the same process.

On the other hand, it can be seen that the coupling constant $\lambda^\prime$ can be measured with 4\% accuracy using the threshold production of top partner pair production. In this case, the center of mass energy required for the measurement is smaller than that of the higgs associated production. Furthermore, using the threshold production of top quark and top partner, the coupling constant $\lambda^\prime$ can be measured with 20\% accuracy even if the center of mass energy is around 500 GeV.

\section{Summary}
\label{sec: summary}

We have studied the capability of future collider experiments to test the little higgs mechanism. The mechanism predicts a certain relation between coupling constants in the top sector of the little higgs scenario. The test of the mechanism is essentially equivalent to the measurement of the Yukawa coupling of the top partner, namely the coupling constant $\lambda^\prime$. It is therefore very important to investigate how accurately $\lambda^\prime$ can be measured at the experiments. With the use of an appropriate representative point, we found that the coupling constant can be measured with a few percent accuracy using the higgs associated productions ($e^+e^- \to T \bar{T} h$) and the threshold production ($e^+e^- \to T \bar{T}$), when the center of mass energy is ${\cal O}(1)$ TeV. On the other hand, using the threshold production ($T\bar{t} + t\bar{T}$), the measurement of $\lambda^\prime$ is still possible with a few ten percent accuracy even if the center of mass energy is around 500 GeV. When the top partner is discovered at the LHC experiment, the processes emphasized in this article will be important to confirm whether the top partner is really the one predicted by the little higgs scenario or not.

\section*{Acknowledgments}

This work is supported by Grant-in-Aid for Scientific research from the Ministry of Education, Science, Sports, and Culture (MEXT), Japan (Nos. 23740169, 22244031, 22244021 for S.M.), and also by World Premier International Research Center Initiative (WPI Initiative), MEXT, Japan.

\appendix

\section{Effective lagrangian}
\label{app: derivation}
\subsection{Derivation}
\begin{table}[t]
\centering
\begin{tabular}{|c|llll|}
\hline
Dim. & \multicolumn{4}{c|}{Operators} \\
\hline
3 & $\bar{U}_L U_R$ & $\bar{U}_L u_{3R}$ & & \rule{0ex}{2.5ex} \\
4 & $\bar{U}_L \slashed{D} U_L$ & $\bar{U}_R \slashed{D} U_R$ & $\bar{Q}_{3L} H^c U_R$ &
\rule{0ex}{2.5ex} \\
5
& $\bar{U}_L U_R |H|^2$
& $\bar{U}_L (H^c)^\dagger \slashed{D} Q_{3L}$
& $\bar{U}_L \sigma^{\mu\nu} u_{3R} B_{\mu\nu}$
& $\bar{U}_L \sigma^{\mu\nu} (\lambda^a/2) u_{3R} G_{\mu\nu}^a$
\rule{0ex}{2.5ex} \\
& $\bar{U}_L u_{3R} |H|^2$
& $\bar{Q}_{3L} H^c \slashed{D} U_L$
& $\bar{U}_L \sigma^{\mu\nu} U_R B_{\mu\nu}$
& $\bar{U}_L \sigma^{\mu\nu} (\lambda^a/2) U_R G_{\mu\nu}^a$
\rule{0ex}{2.5ex} \\
\hline
\end{tabular}
\caption{\small Lorentz and gauge invariant operators including $U_L$ and $U_R$.}
\label{tab: operators}
\end{table}

We first consider operators which are invariant under Lorentz and gauge symmetries of the SM. The operators including top partners $U_L$ and $U_R$ are shown in Table~\ref{tab: operators} up to dimension five, where $\slashed{D} = D_\mu \gamma^\mu$ with $D_\mu$ being the covariant derivative, $G_{\mu\nu}^a (B_{\mu\nu})$ is the field strength tensor of gluon (hyper-charge gouge boson), $\sigma_{\mu\nu} = i[\gamma_\mu, \gamma_\nu]/2$, and $\lambda^a$ is Gell-Mann matrix. Hermitian conjugates of the operators are not shown. Since the quantum number of $u_{3R}$ is exactly the same as that of $U_R$, the bilinear operator $\bar{U}_L u_{3R}$ can be eliminated by an appropriate redefinition of $U_R$. Dimension five operators $\bar{U}_L (H^c)^\dagger \slashed{D} Q_{3L}$ and $\bar{Q}_{3L} H^c \slashed{D} U_L$ are also eliminated by using the equations of motion. On the other hand, dipole-type operators $\bar{U}_L \sigma^{\mu\nu} u_{3R} B_{\mu\nu}$, $\bar{U}_L \sigma^{\mu\nu} (\lambda^a/2) u_{3R} G_{\mu\nu}^a$, $\bar{U}_L \sigma^{\mu\nu} U_R B_{\mu\nu}$ and $\bar{U}_L \sigma^{\mu\nu} (\lambda^a/2) U_R G_{\mu\nu}^a$ do not contain higgs boson, so that they are suppressed by the cutoff scale $\Lambda \simeq$ 10 TeV. Furthermore, they are expected to appear radiatively from UV-completion theory of the little higgs scenario, and coefficients in front of the operators will be suppressed by 1-loop factor. The dipole-type operators are therefore negligible compared to other dimension five operators. The effective lagrangian including top partners is therefore given by
\begin{eqnarray}
{\cal L}
&=&
{\cal L}_{\rm SM}
+ \bar{U}_L i\slashed{D} U_L
+ \bar{U}_R i\slashed{D} U_R
- (m_U \bar{U}_L U_R + h.c.)
\nonumber \\
&-& y_U \bar{Q}_{3L} H^c U_R
- (\lambda^\prime/m_U) \bar{U}_L U_R |H|^2
- (\lambda/m_U) \bar{U}_L u_{3R} |H|^2
+ h.c.,
\label{eq: effective lagrangian app}
\end{eqnarray}
where ${\cal L}_{\rm SM}$ is the SM lagrangian. Parameters $m_U$, $y_U$, $\lambda$, and $y_3$ (top yukawa in ${\cal L}_{\rm SM}$) are taken to be real by appropriate redefinitions of $Q_{3L}$, $u_{3L}$, $U_L$, and $U_R$.

Using the effective lagrangian ${\cal L}$, quadratically divergent corrections to the higgs mass term from the top sector of the little higgs scenario turn out to be
\begin{eqnarray}
\delta \mu^2_t
\simeq
\left(
\frac{3 y_3^2}{4\pi^2} + \frac{3 y_U^2}{4\pi^2} + \frac{3 \lambda^\prime}{2\pi^2}
\right) \Lambda^2.
\end{eqnarray}
Since the corrections should be vanished at 1-loop level because of the little higgs mechanism, we have a special relation among the parameters, $-2\lambda^\prime = y_3^2 + y_U^2$.

\subsection{Correspondence to specific little higgs models}
For convenience we show correspondence between the effective lagrangian
(\ref{eq: effective lagrangian app}) and
specific little higgs models written by a non-linear sigma model.

{\subsubsection*{The littlest higgs model }
The littlest higgs model is well-studied \cite{Perelstein:2003wd}, so it is worth
showing the correspondence of the effective lagrangian. This models is
embeds the electroweak sector of the standard model in an $SU(5)/SO(5)$
non-linear sigma model. The global $SU(5)$ is broken by gauging an
$[SU(2)\times U(1)]^2$ subgroup. So 4 of 14 Nambu-Goldstone bosons (NGBs) are absorbed by the
broken gauge symmetry, and there are 10 NGBs.  
From a non-linear sigma model of the littlest higgs model, its top quark
sector is given by
\begin{eqnarray}
{\cal L}_{top}^{\rm littlest} = -\lambda_1 f \bar{U}_L u_{3R}^{\prime}-\lambda_2 f \bar{U}_L U_{R}^{\prime} 
+\sqrt{2} \lambda_1 \bar{Q}_{3L} H^c u_{3R}^{\prime}  \nonumber\\
+\frac{\lambda_1}{f}\bar{U}_L u_{3R}^{\prime} |H|^2 +h.c. \mbox{ }, 
\label{LittlestHiggs-1}
\end{eqnarray}
where  $f$ is an energy scale at which $SU(5)$ is spontanously broken down to
$SO(5)$. In order to remove one bilinear term, we define $u_{3R}$ and $U_R$ which are given by 
\begin{eqnarray}
 \left(\begin{array}{c}u_{3R} \\ U_{R}\end{array}\right)=
\left(\begin{array}{cc}\frac{\lambda_2}{\sqrt{\lambda_1^2+\lambda_2^2}} & \frac{-\lambda_1}{\sqrt{\lambda_1^2+\lambda_2^2}} \\\frac{\lambda_1}{\sqrt{\lambda_1^2+\lambda_2^2}} & \frac{\lambda_2}{\sqrt{\lambda_1^2+\lambda_2^2}}\end{array}\right)
  \left(\begin{array}{c}u_{3R}^{\prime} \\ U_{R}^{\prime}\end{array}\right),
\end{eqnarray}
and then
(\ref{LittlestHiggs-1}) becomes
\begin{eqnarray}
{\cal L}_{top}^{\rm littlest} = -\hat{\lambda} f \bar{U}_L U_{R} 
+\frac{\sqrt{2} \lambda_1 \lambda_2}{\hat{\lambda}} \bar{Q}_{3L} H^c u_{3R} 
+\frac{\sqrt{2} \lambda_1^2 }{\hat{\lambda}} \bar{Q}_{3L} H^c U_{R} \nonumber\\
+\frac{\lambda_1 \lambda_2}{\hat{\lambda} f}\bar{U}_L u_{3R} |H|^2 
+\frac{\lambda_1^2 }{\hat{\lambda} f}\bar{U}_L U_{R} |H|^2 
+h.c. \mbox{  },
\label{LittlestHiggs-2}
\end{eqnarray}
where $\hat{\lambda}=\sqrt{\lambda_1^2+\lambda_2^2}$. We compare
(\ref{LittlestHiggs-2}) to (\ref{eq: effective lagrangian}) and 
obtain following relations:
\begin{eqnarray}
 \begin{array}{c}m_{U}= \hat{\lambda}f \\\end{array}, 
  \begin{array}{c}y_3=-\sqrt{2}\lambda_1\lambda_2/\hat{\lambda}\\ y_{U}=-\sqrt{2}\lambda_1^2/\hat{\lambda}\end{array},
 \begin{array}{c}\lambda=-\lambda_1\lambda_2 \\ \lambda^{\prime}=\lambda_1^2\end{array} \mbox{  }.
\end{eqnarray}
These coefficients actually hold the relation (\ref{eq: LH relation}).

\subsubsection*{The simplest lttle higgs model} 
We also consider the simplest little higgs model \cite{Kaplan:2003uc,Schmaltz:2004de}, which has an
$[SU(3)\times U(1)]^2$ global symmetry spontaneously broken down to an
$[SU(2)\times U(1)]^2$ subgroup by two vacuum expectation values which are
$f_1$ and $f_2$. The diagonal subgroup of $[SU(3)\times U(1)]^2$ is
gauged, and 5 of 10 NGBs are absorbed by broken gauge
fields. The remaining NGBs contains the higgs boson. 
The top quark sector of this model from a non-linear sigma model is given by 
\begin{eqnarray}
{\cal L}_{top}^{\rm simple} = 
-\lambda_1 f_1 \bar{U}_L U_{R1}
- \frac{\lambda_1 f_2}{f} \bar{Q}_{3L} H^c U_{R1} 
+\frac{\lambda_1 f_2^2}{2f_1 f^2} \bar{U}_L U_{R1} |H|^2 \nonumber\\
-\lambda_2 f_2 \bar{U}_L U_{R2}
+\frac{\lambda_2 f_1}{f}  \bar{Q}_{3L} H^c U_{R2} 
+\frac{\lambda_2 f_1^2}{2f_2 f^2} \bar{U}_L U_{R2} |H|^2 +h.c. \mbox{ }, 
\end{eqnarray}
where $f=\sqrt{f_1^2+f_2^2}$.
Redefine the right-handed top quarks as in the previous model in the following way
\begin{eqnarray}
 \left(\begin{array}{c}U_{R} \\ u_{3R}\end{array}\right)=
\left(\begin{array}{cc}\frac{\lambda_1 f_1}{M} & \frac{\lambda_2 f_2}{M}
      \\\frac{-\lambda_2 f_2}{M} & \frac{\lambda_1 f_1}{M}\end{array}\right)
  \left(\begin{array}{c}U_{R1} \\ U_{R2}\end{array}\right),
\end{eqnarray}
where $M= \sqrt{\lambda_1^2 f_1^2 +\lambda_2^2 f_2^2}$. Then the top sector becomes
\begin{eqnarray}
{\cal L}_{top}^{\rm simple} 
&=& -M \bar{U}_L U_{R} \nonumber\\
&+&\frac{\lambda_1\lambda_2 (f_1^2+f_2^2) }{fM} \bar{Q}_{3L} H^c u_{3R}
-\frac{ f_1f_2(\lambda_1^2 -\lambda_2^2)}{fM} \bar{Q}_{3L} H^c U_{R}  \nonumber\\
&+&\frac{\lambda_1\lambda_2 }{2f^2 M}\left( \frac{f_1^3}{f_2}-\frac{f_2^3}{f_1}\right)\bar{U}_L u_{3R} |H|^2 
+\frac{\lambda_1^2 f_2^2 +\lambda_2^2 f_1^2}{2 f^2M}\bar{U}_L U_{R} |H|^2  \nonumber\\
&+&h.c. \mbox{}
\end{eqnarray}
As a result, we have 
\begin{eqnarray}
 \begin{array}{c}m_{U}= M \\\end{array}, 
  \begin{array}{c}y_3=-\frac{\lambda_1\lambda_2 (f_1^2+f_2^2) }{fM}\\ 
y_{U}=\frac{ f_1f_2(\lambda_1^2 -\lambda_2^2)}{fM} \end{array},
 \begin{array}{c}\lambda=-\frac{\lambda_1\lambda_2 }{2f^2}\left(\frac{f_1^3}{f_2}-\frac{f_2^3}{f_1}\right) \\ 
\lambda^{\prime}=- \frac{\lambda_1^2 f_2^2 +\lambda_2^2 f_1^2}{2 f^2}\end{array} \mbox{  }.
\end{eqnarray}
These also satisfy the relation (\ref{eq: LH relation}).

\section{Representative point}
\label{app: representative point}

The representative point used in our analysis has been chosen so that it is consistent with current experimental data and also attractive from the viewpoint of the naturalness to keep the higgs mass ($m_h$) at the electroweak scale. The most stringent constraints on the parameter $\sin \theta_L$ comes from the electroweak precession tests. Contributions to $S$, $T$, and $U$ parameters~\cite{Peskin:1991sw} from the top partner have already been calculated in Ref.~\cite{Hubisz:2005tx}. On the other hand, hadron collider experiments give the lower bound on the mass of top partner ($m_T$). Currently, the CDF collaboration at the Tevatron experiment gives the stringent bound on $m_T$ thorough the semi-leptonic mode of the process, $p \bar{p} \to T \bar{T} \to b b W W~$\cite{TT}. The upper bound on $m_T$ is obtained from the viewpoint of the naturalness. Though quadratically divergent corrections to $m_h$ are cancelled at 1-loop level, logarithmic corrections still remain at this level. Too large contributions therefore cause a fine-tuning to keep $m_h$ at the electroweak scale~\cite{Perelstein:2003wd}. Logarithmic contributions from the top sector are given by
\begin{eqnarray}
\delta \mu_t^2
=
y_t^2 \frac{3 m_U^2}{8\pi^2} \ln \left(\frac{\Lambda^2}{m_U^2}\right)
\simeq
\frac{3 m_T^2}{4 \pi^2}
\frac{m_t^2}{v^2}
\ln \left(\frac{\Lambda^2}{m_T^2}\right),
\end{eqnarray}
with imposing the relation in eq.(\ref{eq: LH relation}). Here, $\Lambda \simeq$ 10 TeV is the cutoff scale of the little higgs scenario. The level of the fine-tuning is parameterized by $F \equiv m_h^2/(2 \delta \mu_t^2)$.

\begin{figure}[t]
\begin{center}
\includegraphics[scale=0.26]{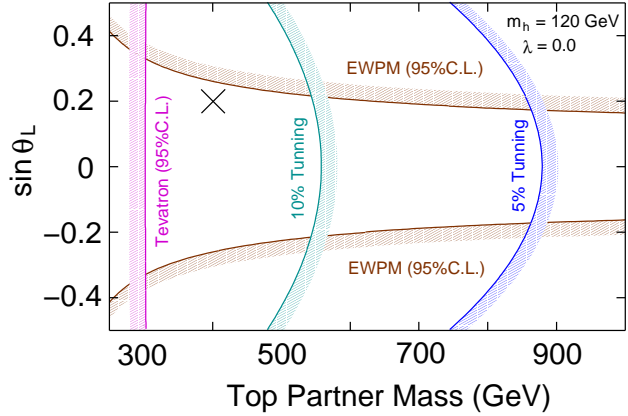} \\
\caption{\small Constraints on the model parameters and the representative point.}
\label{fig: representative point}
\end{center}
\end{figure}

These constraints on the $(m_T, \sin \theta_L)$-plane are summarized in Fig.~\ref{fig: representative point}, where $\lambda^\prime$ is set so that it satisfies the little higgs relation in eq.(\ref{eq: LH relation}). The representative point is denoted by the cross. It can be seen that the point satisfies the experimental constraints and the level of fine-tuning is less than 10\% ($F \leq 0.1$).

\section{Non-relativistic lagrangian}
\label{app: threshold productions}

Derivation of the non-relativistic lagrangian for threshold productions is considered in this appendix. For concrete discussion, we consider the process of top partner pair production at the threshold energy $\sqrt{s} \simeq 2m_T$. With the use of the effective lagrangian in eq.(\ref{eq: effective lagrangian app}), interactions relevant to the threshold production are
\begin{eqnarray}
{\cal L}
&=&
\bar{T}(i\slashed{\partial} - m_T)T
-\frac{1}{2}h(\partial^2 + m_h^2)h
+\frac{1}{2}G_\mu^a \partial^2 G^{\mu a}
+\frac{1}{2} A_\mu \partial^2 A^\mu
\nonumber \\
&&
+\bar{e}i\slashed{\partial}e
-y_T\bar{T}Th
-g_s\bar{T}\slashed{G}T
-Q_T\bar{T}\slashed{A}T
-v_T\bar{T}\slashed{Z}T
\nonumber\\
&&
+\frac{1}{2}Z_\mu
\left[ g^{\mu\nu}(\partial^2 + m_Z^2) - \partial^\mu \partial^\nu \right] Z_\nu
-Q_e \bar{e}\slashed{A}e
-\bar{e}\slashed{Z}(v_e + a_e\gamma^5)e,
\end{eqnarray}
where $e$ is the electron field. Unitarity (Feynman) gauge is adopted for $Z$ boson (photon and gluon). After integrating higgs and gluon fields out from the lagrangian and expanding the top partner field $T$ in terms of its velocity (non-relativistic expansion), namely, $T \simeq (e^{-im_Tt} \eta + ie^{im_Tt} \nabla \cdot \sigma \chi/2m_T,~e^{im_Tt} \chi - ie^{-im_Tt} \nabla \cdot \sigma \eta/2m_T)^T$, the effective action for non-relativistic fields $\eta$ and $\chi$ is obtained as
\begin{eqnarray}
S_{\rm eff}
&=&
\int d^4x
\left[
\eta^\dagger
\left(i\partial_t + \frac{\nabla^2}{2m_T} + i\frac{\Gamma_T}{2} \right) \eta
+ \chi^\dag
\left(i\partial_t - \frac{\nabla^2}{2m_T} - i\frac{\Gamma_T}{2} \right) \chi
\right]
\nonumber \\
&+&
\int d^4x~d^4y~\delta(x^0-y^0)
\frac{V_{TT}({\bf x}-{\bf y})}{6}
\left[\eta^\dagger(x) \sigma^i \chi(y)\right]
\left[\chi^\dagger(y) \sigma^i \eta(x)\right]
\nonumber \\
&+&
\int d^4x
\left[
  Q_T A^i (e^{2im_T} \eta^\dagger \sigma^i \chi + h.c.)
+ v_T Z^i (e^{2im_T} \eta^\dagger \sigma^i \chi + h.c.)
\right]
\nonumber \\
&+&
\int d^4x
\left[
\bar{e} i\slashed{\partial} e
-Q_e \bar{e} \slashed{A} e
-\bar{e} \slashed{Z} (v_e + a_e\gamma^5) e
\right],
\label{eq: NR effective action}
\end{eqnarray}
where the superscript '$i$' runs from one to three, $\sigma^i$ is the Pauli matrix, and $\Gamma_T$ is the total decay width of the top partner. The potential $V_{TT}({\bf r})$ has already been defined in eq.(\ref{eq: potential V_TT}). In addition to the terms in above action (\ref{eq: NR effective action}), we also have another potential terms such as the one composing spin-zero state ($\propto [\eta^\dagger(x) \chi(y)]$) and those composing color octet states ($\propto [\eta^\dagger(x) \lambda^a \sigma^i \chi(y)]$ \& $[\eta^\dagger(x) \lambda^a \chi(y)]$), where $\lambda^a$ is the Gell-Mann matrix. Since these states are never produced at high energy $e^\pm$ collisions, we have dropped these terms from the effective action. In order to calculate the cross section at the threshold energy, it is convenient to introduce the auxiliary fields $\sigma^i_t$ and $\sigma^{i\dagger}_t$ by inserting following identities into the action,
\begin{eqnarray}
1 = \int {\cal D}\sigma^i{\cal D}s^{i \dagger}
\exp
\left[
\frac{i}{2} \int d^3x~d^3y~dt~\sigma^i_t({\bf x}, {\bf y})
\left\{
s^{i \dagger}_t({\bf y}, {\bf x})
-
\frac{\eta^\dagger({\bf x}, t) \sigma^i \chi({\bf y}, t)}{\sqrt{3}}
\right\}
\right],
\nonumber \\
1 = \int {\cal D}\sigma^{i \dagger} {\cal D}s^i
\exp
\left[
\frac{i}{2} \int d^3x~d^3y~dt~\sigma^{i \dagger}_t({\bf x}, {\bf y})
\left\{
s^i_t({\bf y}, {\bf x})
-
\frac{\chi^\dagger({\bf y}, t) \sigma^i \eta({\bf x}, t)}{\sqrt{3}}
\right\}
\right].
\end{eqnarray}
Here, we integrate all fields out except the auxiliary fields ($\sigma^i_t$ and $\sigma^{i\dagger}_t$) from the effective action in eq.(\ref{eq: NR effective action}), and canonically normalize $\sigma^i_t$ and $\sigma^{i \dagger}_t$ as
\begin{eqnarray}
\phi_{TT}^i({\bf r}, R)
=
\frac{1}{\sqrt{2}V_{TT}({\bf r})}
\left[
\sigma^i_t({\bf x}, {\bf y})
+
2\sqrt{3} e^{2im_Tt} \delta({\bf x} - {\bf y}) \left\{v_T Z^i(x) + Q_T A^i(x) \right\}
\right],
\end{eqnarray}
where variables ${\bf r}$, ${\bf R}$ and $R^0$ are defined by ${\bf r} = {\bf x} - {\bf y}$ and ${\bf R} = ({\bf x} + {\bf y})/2$ and $R^0 = t$, respectively. We then obtain the non-relativistic lagrangian directly used to calculate the cross section for the threshold production of the top partner pair,
\begin{eqnarray}
S_{\rm NR}^{(TT)}
&=&
\int d^4R~d^3r~
\phi_{TT}^{i \dagger}({\bf r}, R)
\left[
i\partial_{R^0}
+\frac{\nabla^2_{\bf R}}{4m_T}
+\frac{\nabla^2_{\bf r}}{m_TT}
-V_{TT}({\bf r})
+i\Gamma_T
\right] \phi_{TT}^i({\bf r}, R),
\nonumber \\
&-&
\int d^4R
\left[
\sqrt{6} e^{2i m_T R^0}
\phi_{TT}^{i \dagger}({\bf 0}, R)
\left\{ v_T Z^i(R) + Q_T A^i(R) \right\}
+ h.c.
\right]
\nonumber \\
&+&
\int d^4x
\left[
\bar{e} i\slashed{\partial} e
-Q_e \bar{e} \slashed{A} e
-\bar{e} \slashed{Z} (v_e + a_e\gamma^5) e
\right].
\label{eq: NR action}
\end{eqnarray}
The fields $\phi^i_{TT}$ and $\phi^{i \dagger}_{TT}$ are describing the annihilation and creation of top partner pair, respectively. It can be seen that the effect of the threshold singularity is involved in the first term of the action with the form of the Schr$\ddot{\rm o}$dinger equation. The non-relativistic action for the case of top quark and top partner production can be obtained in the same procedure of top partner pair production. 

Cross section formulae for threshold productions are obtained by using the optical theorem, which is related to the calculation of the forward scattering amplitude. The cross section of the top partner pair production is, for instance, given by
\begin{eqnarray}
\sigma_{TT}
= \frac{1}{s}{\rm Im} 
\left[ {\cal M} (e^+ e^- \to T \bar{T} \to e^+ e^-) \right].
\end{eqnarray}
Amplitude ${\cal M}$ is calculated using above non-relativistic action, and eventually obtain the cross section formula in eq.(\ref{eq: sigma_TT}). It is worth notifying that the solution of Schr$\ddot{\rm o}$dinger equation in eq.(\ref{eq: Schrodinger TT}) is required to calculate ${\cal M}$ quantitatively, as already mentioned in section \ref{sec: threshold}. Here, the green function $G_{TT}(E; {\bf r}, {\bf r}^\prime)$ in this equation is nothing but the Fourier transform of the following two-point function,
\begin{eqnarray}
G_{TT}(E, {\bf r}, {\bf r}^\prime)
=
i\int d^4R
\langle 0 | T \left[
\phi_{TT}^i({\bf r}, R)
\phi_{TT}^{i \dagger}({\bf r}^\prime, R^\prime)
\right] | 0 \rangle
e^{i E (R - R^\prime)},
\end{eqnarray}
where $\langle 0 |T[\cdots]| 0 \rangle$ is the vacuum expectation value of the time-ordered product.

\end{document}